\documentclass{amsart}
\usepackage[all]{xy}
\usepackage{hyperref}
\hypersetup{
  colorlinks = true,
  linkcolor=red,   % color of internal links
  citecolor=blue,   % color of links to bibliography
  urlcolor=blue,    % color of external links
  pagebackref=true,
  implicit=false,
  bookmarks=true,
  bookmarksopen=true,
  pdfdisplaydoctitle=true
}
\begin{document}

\title[TWO-DIMENSIONAL QUANTUM FIELD MODELS]{TWO-DIMENSIONAL QUANTUM FIELD MODELS\\ (WITH APPLICATIONS TO
  LATTICE STATISTICAL MECHANICS)}

\author{P. FALCO}

\address{Department of Mathematics, California State University\\
  Northridge,
  CA 91330\\
  E-mail: pierluigi.falco@csun.com\\
  www.csun.edu}

\begin{abstract}
  Two dimensional toy models display, in a gentler setting, many
  salient aspects of Quantum Field Theory.  Here 
I discuss a concrete two dimensional case,
  the Thirring model, which
  illustrates several important concepts of this theory: 
  the anomalous dimension
  of the fields; the exact solvability; the anomalies of the
  Ward-Takahashi identities.  Besides,  I give a glimpse of the decisive
  role that this model plays in the study of an apparently
  unrelated topic: correlation critical exponents of 
  two dimensional lattice systems of Statistical
  Mechanics.
\end{abstract}


\maketitle

\section{Introduction}
The Thirring model \cite{Thir58} is a toy model for a
self-interacting 2-dimensional relativistic fermion field. 
An enormous number of articles in the physical 
and mathematical literatures testify its
importance in the study of Quantum Field Theory (QFT). 

In the first part of this review I will deal
with  physicists'
conjectures (Sec. \ref{2.1} and \ref{2.2}) and mathematical results (Sec \ref{2.3}) 
about the two most salient features of this model: the
\emph{bosonization} of the currents; the
\emph{exact solvability} of the massless case. 
In the second part (Sec. \ref{3.0}), I will discuss a third, less
known, aspect: 
the Thirring model is the \emph{scaling limit} of
several lattice models of Statistical Mechanics.  In prospective, 
I consider this  latter feature  as one of the most promising
tool for the study of critical exponents in two dimensional Statistical 
Mechanics.

\section{Thirring Model}
Spacetime is two-dimensional,  ${\bf x}:=(x_0,x_1)$ and, 
following physicists' conventions, a repeated index $\mu$ 
implies a sum over $\mu=0,1$. 
Put $\gamma^0=
\begin{pmatrix}
  0&1\cr1&0
\end{pmatrix}
$ and $\gamma^1=
\begin{pmatrix}
  0&-i\cr i&0
\end{pmatrix}
$; % be the generators of the two-dimensional Euclidean Clifford Algebra;
let $\not\!\partial:=\gamma^\mu \partial^\mu$ 
be the Dirac operator; and let $\bar\psi_{\bf x}=
\begin{pmatrix}
  \bar \psi_{{\bf x},+}\cr \bar\psi_{{\bf x},-}\cr
\end{pmatrix}
$ and $\psi_{\bf x}=(\psi_{{\bf x},+}, \psi_{{\bf x},-})$ 
be two-components anticommuting vector fields.
Given two real parameters,  the coupling constant $\lambda$ and 
the mass $m$, the Euclidean  Thirring
Lagrangian is 
\begin{equation}
\label{tm}
  \int\!\!d^2{\bf x}\; \bar\psi_{\bf x} (i\!\not\!\partial+m) \psi_{\bf
    x} +\frac\lambda 4 \int\!\!d^2{\bf x}\;(\bar\psi_{\bf
    x}\gamma^\mu\psi_{\bf x})(\bar\psi_{\bf
    x}\gamma^\mu\psi_{\bf x})\;.
\end{equation}
Let us consider separately the exact solvability and the bosonization.

\subsection{Massless Thirring model: the formal exact solution }
\label{2.1}
The sequence of early attempts at an exact solution of the massless
model is quite an instructive piece of history of QFT\cite{Wig64}.  At
first, several scholars ``derived'' different sets of explicit
formulas for all the Schwinger functions; however ``\dots\ it was not
clear which of the many partly contradictory equations written by the
various authors were true and which false.''\cite{Wig64}. In this
confusing status, the model fell into disrepute; until
Johnson\cite{Jo61} was finally able to propose formulas for the two
and the four points functions that were  not susceptible to
objections.  Johnson's two points function is
\begin{equation}\label{tpf}
  \tfrac1Z\langle \psi_{{\bf x}} \bar \psi_{0}\rangle =C
  \begin{pmatrix}
    0& \frac{1}{x_0+i x_1}|{\bf x}|^{-\eta} \\\\
    \frac{1}{x_0-i x_1}|{\bf x}|^{-\eta} & 0
  \end{pmatrix}
\end{equation}
where $C$ and $\eta$ are constants; while $Z$ is the \emph{wave
  function renormalization}, i.e. an infinite or zero factor that
one has to divide out in the end of the derivation 
in order to get a finite, non-identically vanishing,
outcome. % In this model, $Z\sim e^{-\eta\cdot\infty}$ and
%$\eta=O(\lambda))$.

Johnson's approach was only partially a derivation; it was mostly a
self-consistency argument. In principle \eqref{tpf}
should be the solution of the integral equation which is obtained by
plugging the Ward-Takahashi Identities
(WTI) into the Schwinger-Dyson Equation (SDE).  In actuality this method has a problem: 
 if $J^\mu_{\bf x}:=\bar
\psi_{\bf x} \gamma^\mu \psi_{\bf x}$ and $J^\mu_{5,{\bf x}}:=\bar
\psi_{\bf x} \gamma^5 \gamma^\mu \psi_{\bf x}$ for
$\gamma^5=i\gamma^0\gamma^1$, the \emph{formal} WTI are
\begin{align}\label{wti}
  \tfrac{1}{Z}\;i\partial^\mu_{\bf z} \langle J^\mu_{\bf z} \psi_{\bf
    x} \bar \psi_{\bf y}\rangle &= \frac{a}{Z}\left[\delta({\bf z}-{\bf
      x})-\delta({\bf z}-{\bf y})\right] \langle\psi_{\bf x} \bar
  \psi_{\bf y}\rangle
  \notag\\
  \tfrac{1}{Z}\;i\partial^\mu_{\bf z} \langle J^\mu_{5,{\bf z}}
  \psi_{\bf x} \bar \psi_{\bf y}\rangle &= \frac{\bar
    a}{Z}\left[\delta({\bf z}-{\bf x})-\delta({\bf z}-{\bf y})\right]
  \gamma^5\langle\psi_{\bf x} \bar \psi_{\bf y}\rangle
\end{align}
with $a=\bar a=1$; but then, plugging them into the SDE, one would obtain
an integral equation that  is solved by \eqref{tpf} with $\eta=0$. 
Namely, one would arrive at the odd 
conclusion that the Thirring model, which is an
interacting field theory, has the same two points function of
the free field theory!  Johnson's idea was to allow for $a\neq1$ and
$\bar a\neq1$, so that, after \emph{formal operations with
  infinities}, he obtained \eqref{tpf}, with
\begin{equation*}
  \eta=\frac{\lambda}{4\pi} (a-\bar a)\;.
\end{equation*}
$a$ and $\bar a$ are then fixed by consistency with the formula
for the four points function: for $\nu=-\bar \nu=
\frac{\lambda}{4\pi}$, 
\begin{equation}
  \label{aba}
  a=\frac{1}{1-\nu}\qquad
  \bar a=\frac{1}{1-\bar \nu}\;.
\end{equation}
Finally, Hagen\cite{Ha67} (by Johnson's method) and
Klaiber\cite{Kla68} (by a different ansatz), found the explicit formulas 
for all the $n$-points
Schwinger functions.\footnote{In fact they also generalized Johnson's
  solution, since they obtained $\nu=\frac{\lambda}{2\pi}(1-\xi)$ and
  $\bar \nu=-\frac{\lambda}{2\pi} \xi$, for any real $\xi$. Several
  years later $\nu$ and $\bar \nu$ would have been called
  Adler-Bell-Jackiw anomalies of the vector and the axial-vector WTI.
  The numerical value of $\xi$ is regularization dependent: 
  for example in the dimensional regularization
  $\xi=1$ and only the axial-vector WTI, i.e. the latter of (\ref{wti}), is anomalous. } The only
non-zero ones are
\begin{equation}
  \label{es}
  \frac1{Z^n}\langle\psi_{{\bf x}_1, \omega_1}\cdots\psi_{{\bf x}_n, \omega_n}
  \bar \psi_{{\bf y}_1, \sigma_1}\cdots\bar \psi_{{\bf y}_n, \sigma_n}\rangle=
  \sum_{\pi}(-1)^\pi G_{\underline \omega, \pi(\underline \sigma)}(\underline {\bf x}, \pi(\underline {\bf y}))
\end{equation}
where $\pi$ runs over the permutations of $n$ elements and $(-1)^\pi$
is the signum of the permutation; and,  for $\omega_j=\pm1$ and
$\sigma_j=\pm1$,
\begin{align}\label{npf}
  G_{\underline \omega, \underline \sigma}(\underline {\bf x},
  \underline {\bf y}) := &\tfrac1Z\langle \psi_{{\bf x}_1,
    \omega_1}\bar \psi_{{\bf y}_1, \sigma_1}\rangle \cdots
  \tfrac1Z\langle \psi_{{\bf x}_n, \omega_n}\bar \psi_{{\bf y}_n,
    \sigma_n}\rangle \notag
  \\
  &\cdot \frac{\prod_{i<j}|{\bf x}_i-{\bf x}_j|^{\eta_{-\omega_i
        \omega_j}} \prod_{i<j}|{\bf y}_i-{\bf y}_j|^{\eta_{-\sigma_i
        \sigma_j}}} {\prod_{i\neq j}|{\bf x}_i-{\bf
      y}_j|^{\eta_{\omega_i \sigma_j}}}
\end{align}
where $\eta_-=\eta$ and $\eta_+$ is a new coefficient. From these
formulas, Wilson\cite{Wil70} derived the 
short-distance behavior of the Schwinger functions  
of local quadratic monomials of the field (i.e. the 
\emph{currents}, see below). 

Is this solution completely satisfactory? As we saw, it
was not obtained by rigorous procedures. Yet Wightman\cite{Wig64} pointed
out that if \eqref{es} satisfied some suitable set of axioms, say the
Osterwalder-Schrader's ones, then one would be entitled to  claim 
that \eqref{es} describes a model of QFT! In our case, surprisingly enough, although
formulas are explicit, it has not been an easy task to verify the
axiom of \emph{reflection positivity}. I will return on this
matter in the section on mathematical results. %  Let me emphasize here
%that this viewpoint is applicable, of course, only in models for which
%there exists an ansatz of exact solution.
  
\subsection{General Thirring model: bosonization}
\label{2.2}
Let us consider, now, two different models. The first is the Thirring
model with a possible mass term (\ref{tm}).  The second is a boson system, called
\emph{sine-Gordon} model: for two real parameters $\beta>0$ and $z$,
the Lagrangian is
\begin{equation*}
  \frac{1}{2\beta}\int\!\!d^2{\bf x}\; (\partial^\mu \phi_{\bf x})^2+ z
  \int\!\!d^2{\bf x}\;:\!\cos \phi_{\bf x}\!:\
\end{equation*}
where $:\!\cos \phi_{\bf x}\!:$ is the \emph{normal ordering} of $\cos
\phi_{\bf x}$ (which corresponds to multiply $ \cos \phi_{\bf x} $ by
an infinite factor).  Coleman\cite{Col75} discovered a surprising
relation between the two models. Define the fermion currents
\begin{equation*}
  J^\mu_{\bf x}:=\bar \psi_{{\bf x}}\gamma^\mu \psi_{{\bf x}}\qquad
  O^\sigma_{\bf x}:=\bar \psi_{{\bf x}}(1+\sigma \gamma^5) \psi_{{\bf
      x}}
\end{equation*}  
(for $\mu=0,1$ and $\sigma=\pm1$); and define the boson observables
\begin{equation*}
  \mathcal{J}^\mu_{\bf
    x}:=-\frac{1}{\sqrt\pi}\epsilon^{\mu\nu}\partial^\nu \phi_{{\bf
      x}}\qquad \mathcal{O}^\sigma_{\bf x}:=:\!e^{i\sigma \phi_{\bf
      x}}\!:
\end{equation*}
where $:\!e^{i\sigma \phi_{\bf x}}\!:$ equals $e^{i\sigma \phi_{\bf
    x}}$ times an infinite factor.  The bosonization is the claim that
there exists a choice of $\beta$ and $z$ as function of $\lambda$ and
$m$, with $\beta=4\pi+O(\lambda)$ and $z=O(m)$, such that, if ${\bf
  x}_1, \ldots, {\bf x}_n, {\bf y}_1, \dots, {\bf y}_m$ are two by two
different,
\begin{equation*}
  \zeta_J^n \zeta_O^m\; \langle J^{\mu_1}_{{\bf x}_1}\cdots
  J^{\mu_n}_{{\bf x}_n} O^{\sigma_1}_{{\bf y}_1}\cdots
  O^{\sigma_m}_{{\bf y}_m}\rangle_{Th} = \langle
  \mathcal{J}^{\mu_1}_{{\bf x}_1}\cdots \mathcal{J}^{\mu_n}_{{\bf x}_n}
  \mathcal{O}^{\sigma_1}_{{\bf y}_1}\cdots \mathcal{O}^{\sigma_m}_{{\bf
      y}_m}\rangle_{sG}
\end{equation*}
where $\zeta_J$ and $\zeta_O$ are possibly infinite or zero prefactor
that are necessary to have finite, non-zero correlations of local products of
fermions; and $\langle \cdot\rangle_{Th}$ and $\langle
\cdot\rangle_{sG}$ are the Thirring and the sine-Gordon expectations
in the sense of path-integrals. Note two facts. First, for $m=0$ one
has $z=0$, therefore the current correlations of the massless Thirring
model are \emph{free boson correlations}. Second, for $m\neq0$ the
correlations of the Thirring model are expected to decay exponentially;
and so are the boson correlations, then, even though the
sine-Gordon model does not have a mass term in the Lagrangian: this
suggests a \emph{dynamically generated mass} for the boson field.
Finally let me remark that the bosonization at $m=0$ was known much
earlier in Condensed Matter Physics\cite{Tom50, MaLi65}; besides,  the
general Coleman's construction was later made more precise by
Mandelstam\cite{Man75}.

\subsection{Some Mathematical Results on the Thirring Model}
\label{2.3}
As one might expect at this point of the reading, the first mathematical results 
on the Thirring model used the exact solvability or the bosonization as \emph{tools} 
for constructing the model. In the massless case, one approach was to prove 
that  Hagen's and Klaiber's formulas are  reflection positive (the 
other Osterwalder-Schrader's axioms are clearly fulfilled)\cite{CRW85};
another approach was to obtain the Thirring model via a limiting 
procedure from  the Luttinger model, a model of Condensed Matter Theory
that had been exactly solved earlier by bosonization \cite{MaLi65}. 
A survey of these results is in\cite{CaLa02}. 
In regard to the massive case, bosonization is again the key idea, because the
\emph{spectrum} and the 
\emph{scattering matrix} of the sine-Gordon model can be
exactly computed; from that, it is expected that the 
asymptotic behavior of the fermion Schwinger functions can be reconstructed.
For a  survey  the reader can consult the contribution 
of G. Niccoli to these proceedings \cite{Ni12}. 

Here we do not discuss further these methods.  
They are certainly of great mathematical interest.  
However, these methods are spoiled by physically irrelevant  modifications of the model, 
such as: replacement of the continuum spacetime with a lattice; 
addition of  interacting terms of order higher than four; small deformation 
of the linear dispersion; and others. Here we want to focus instead
on approaches that are relevant for the application that 
we will emphasize in the next section: the  study of the scaling limit of 
lattice models of Statistical Mechanics. 

The main idea of such alternative 
viewpoint is a rigorous reformulation of the physicists' Renormalization Group (RG). 
A first result along this line was \cite{FrSe76}. It contained the proof of
analyticity in $z$ of the sine-Gordon model for $|z|$ small enough and
$\beta\in [0,4\pi)$.  This was an important achievement, for
Coleman's conjecture was based on the identity of the coefficients of
the perturbative expansion of the Thirring and of the sine-Gordon
models. But it was not a \emph{proof} of bosonization, because the
authors could \emph{not} deal with the perturbation theory of the
Thirring model as well.  
%Their approach is based on the
%Glimm-Jaffe-Spencer renormalization group (RG) method\cite{GlJaSp73},
%which was the first rigorous implementation of the Wilson's ideas -
%and the milestone in the subject - but had the limitation of being
%applicable only if the model is \emph{super-renormalizable}. This is
%the case of the sine-Gordon model, for $\beta\in [0,4\pi)$; but it is
%not the case of the Thirring model, which is instead
%\emph{renormalizable} and so requires a more sophisticated treatment.
(Besides, to avoid the mathematical difficulty of the spontaneous mass
generation, in \cite{FrSe76} a mass term $\mu^2\int\!d^2{\bf x}
\;\phi_{\bf x}^2$ is built in the Lagrangian of the sine-Gordon model,
which then corresponds to the bosonization of a slightly different
fermion system, called Schwinger-Thirring model). After some years,
Dimock\cite{Dim98}, using the RG approach of Brydges and
Yau\cite{BY90}, extended the result of\cite{FrSe76} to $\beta\in [0,
\frac{16}{3}\pi)$.  (He avoided the problem of the spontaneous mass
generation by confining the boson interaction term to a finite
volume).  However, he still had no result for the Thirring model.

Some of the problems that were left open by these early papers were
settled in a series of results in collaborations with G. Benfatto and
V. Mastropietro. Starting from the path-integral formulation of the
Thirring model, regularized by the presence of an ultraviolet and an
infrared cutoffs, we proved the following
facts\cite{Fa06,BFM07,BFM09}.
\begin{enumerate}
\item\emph{Massless case.} For $|\lambda|$ small enough, there exists
  the limit of removed cutoffs of all the $n$-points Schwinger
  functions. Such limiting functions satisfy the Osterwalder-Schrader
  axioms, and coincide with  Hagen's and Klaiber's ansatz
  \eqref{es}. Besides, bosonization is proved for $|\beta-4\pi|$ small
  enough.
\item\emph{Massive case.} For any $m$ and $|\lambda|$ small enough,
  again there exists the limit of removed cutoffs of all the $n$-points
  Schwinger functions. The Osterwalder-Schrader axioms are
  fulfilled. Besides, bosonization is proved for $|\beta-4\pi|$ small
  enough \emph{if} both the fermion mass term and the boson
  interaction term are \emph{confined to a finite volume} (to overcome
  the difficulty of the spontaneous mass generation).
\end{enumerate}
The major problem that these works leave open, then, is the proof of
bosonization when the spontaneous mass generation is not
artificially avoided. On the other hand, to the best of my knowledge, these are the first
rigorous results for the {\it Schwinger functions} of the  massive Thirring model. 
And for the massless
Thirring model, this was the first time that the exact solution is
\emph{derived from}  (as opposed to 
\emph{assumed in place of}) the regularized path-integral formulation.  
The general approach of these works is the RG technique
developed by Gallavotti and Nicol\'o\cite{GaNi85, Ga85}, and a
technical result called \emph{vanishing of the beta
  function}\cite{BGPS94,BeMa05}. To understand the difficulty of the
problem, let me spend a few more  words on some aspects of these results. From the second
line of \eqref{npf} we see that, in the $n$-points functions, the decay
for large separation of the points is not the same as in the products
of two points functions. Besides, the two points function itself does
not have the same large distance decay of the free field, but displays
an \emph{anomalous exponent}, $\eta$. In the jargon of the
Renormalization Group, the Thirring model is said to describe a
\emph{Non-Gaussian Fixed Point} (in fact, since the value of $\eta$
depends upon $\lambda$, here we have a case of an \emph{interval} of
non-Gaussian fixed points).

Let me also emphasize some differences w.r.t.  Johnson's work
(see\cite{BFM06} for more details). (i) Of course, we do not work with
infinities, but the theory is regularized and the cutoffs are removed
in the final results only.  At the same time, we do not modify by
hand the formal WTI: in our scheme  the coefficients $\nu\neq0$ and $\bar \nu\neq0$
naturally arise in the limit of removed cutoffs from terms that, in
formal treatments, are considered negligible.  That is known to
physicists as Adler-Bell-Jackiw mechanism\cite{Ad69,BeJa69}.  (ii) Our
$\nu$ and $\bar\nu$ are not linear in $\lambda$; besides, our value
for $\frac{\eta}{a-\bar a}$ is not linear in $\lambda$
either\footnote{interesting enough, this fact is due to a ``new
  anomaly'' which is closely related to the exact solvability of the
  model whenever a ``local'' regularization is employed\cite{Fa06}}.
Point (ii) does not mean that Johnson's solution is wrong. It is a
general expectation in QFT that macroscopic quantities, such as
$\eta$, are related to the bare parameters in the Lagrangian, such as
$\lambda$, in a regularization-dependent way.  Indeed, as
counter-proof, a different regularization of the Thirring model, a
``non-local'' one,  does give $\nu$, $\bar \nu$ and $\frac{\eta}{a-\bar
  a}$ linear in $\lambda$\cite{Ma07a}.  That is known to physicists as
Adler-Bardeen theorem\cite{AdBa69}.

Our RG approach has other applications in QFT. In two cases, exact
solutions, previously conjectured by physicists, are rigorously
derived from the regularized path-integral formulation of the problems:
for the Thirring-Wess model (i.e. a fermion field interacting with a
vector model), see\cite{Fa10}; and for a two colors generalization of
the massless Thirring model (which can include also an interaction
that is not rotational invariant), see\cite{BFM11}.  For lack of
space, I do not provide  details here.

\section{Lattice models of Statistical Mechanics}
\label{3.0}
The most basic lattice model of two-dimensional Statistical Mechanics
is the square lattice Ising model with finite range interactions.  In
particular, let us consider the following Hamiltonian: for real $J$
and $K$, and spin $\sigma_{\bf x}=\pm1$,
\begin{equation*}
  H(\underline \sigma) :=-J\sum_{\substack{{\bf x},{\bf x}'\\
    \text{n.n.}}}\sigma_{{\bf x}} \sigma_{{\bf x}'} 
  -K\sum_{\substack{{\bf x},{\bf
      x}'\\ \text{n.n.n.}}}\sigma_{{\bf x}} \sigma_{{\bf x}'}
\end{equation*}
where: the first sum is over nearest-neighbor (n.n.) sites; the
second is over next nearest-neighbor (n.n.n.) sites. The case $K=0$ is
the celebrated one, the one for which Onsager discovered the
non-trivial exact formula of the free energy\cite{Ons44}. On the contrary,
providing any rigorous result for $K\neq 0$ remained for many years
an open problem;  until very recently, when Spencer\cite{Spe00}
proposed how to \emph{rigorously} calculate certain critical exponents
for $K\neq0$.  To understand his idea, it is important to mention that
the reason behind the exact solvability at $K=0$ is the equivalence of
the Ising model with a system of \emph{free} lattice
fermions\cite{KaOn49,Kas63,SML64}.  Spencer's suggestion is to use the
same fermion re-phrasing also in the non-solvable case.  Of course the
lattice fermion field, this time, is not free, but self-interacting;
yet, handled by RG, the self-interaction turns out to be an
\emph{irrelevant} perturbation of the fermion free field. As a consequence, if
$|K|$ is small enough w.r.t. $|J|$, critical exponents should
remain unchanged.  Guided  by  these ideas, Pinson and Spencer proved
the following result\cite{PS00}. The \emph{local energy} random variable is  
\begin{equation*}
  O_{\bf x}:= \sum_{\substack{{\bf x}'\\ \text{n.n. of }{\bf x}}} \sigma_{\bf x}
  \sigma_{{\bf x}'}\;;
\end{equation*}
then, if $|K|$ is small enough w.r.t. $|J|$, there exists one ($J$ and
$K$ dependent) critical temperature at which,  for large
$|{\bf x}|$,
\begin{equation*}
  \langle O_{\bf x} O_{0}\rangle = \frac{C}{|{\bf x}|^{2\kappa_+}}+o(1)
  \qquad\text{with } \kappa_+=1\;.
\end{equation*}
The fact that the \emph{energy critical exponent}, $\kappa_+$, is
independent of $K$,  hence coincides with the one of the n.n. Ising model,
is a property called \emph{universality}.

The method of the fermion equivalence, which I will call
\emph{interacting fermions picture} (IFP), has a much wider
applicability.  A most natural generalization of the model
is the class of systems made of two (apriori independent) n.n.  Ising models that are
connected to each others by a quartic interaction. The 
Hamiltonian is
\begin{equation*}
  H(\underline \sigma, \underline \tau):= 
  -J\sum_{\substack{{\bf x},{\bf
        x}'\\\text{n.n.}}}\sigma_{\bf x} \sigma_{{\bf x}'} 
  -J\sum_{\substack{{\bf
        x},{\bf x}'\\\text{n.n.}}}\tau_{\bf x} \tau_{{\bf x}'}
  +K\sum_{\substack{{\bf x},{\bf x}'\\\text{n.n.}}} 
  \sum_{\substack{{\bf y},{\bf
        y}'\\\text{n.n.}}}\sigma_{\bf x} \sigma_{{\bf x}'} v({\bf x}-
  {\bf y}) \tau_{\bf y} \tau_{{\bf y}'}
\end{equation*}
where: $\sigma_{\bf x}=\pm1$ and $\tau_{\bf x}=\pm1$ are two spins
located at the same site; $|v({\bf x})|\le C e^{-c |{\bf x}|}$.
This class of \emph{double Ising models} (DIM) does not have just a
mere academic interest, since it encompasses two lattice systems that
are famous for historical and technical reasons: the Ashkin-Teller and
the Eight Vertex models\footnote{an exact formula for the free energy
  of this two models (but not, in general, for all the DIM) is
  available\cite{Bax82}. Though, to avoid confusion, I will not call
  them exactly solvable models for, as opposed to what happens for the
  n.n. Ising model, no exact formula is known for correlation
  functions of local bulk observables.}. 
Of course at $K=0$, called \emph{free fermion point}, 
the model is again exactly solvable. And for $K\neq0$ one can derive
an equivalence with self-interacting lattice fermions. The novelty
w.r.t. the previous case is that the self-interaction is
\emph{marginal}, and so it does change the large distance decay of the
observables. The following facts were proved in \cite{Ma04}.  Define
the \emph{energy} and the \emph{crossover} random variables  to be
\begin{equation*}
  O^+_{\bf x}:= \sum_{\substack{{\bf x}'\\ \text{n.n. of }{\bf x}}} \sigma_{\bf
    x} \sigma_{{\bf x}'} 
  + \sum_{\substack{{\bf x}'\\ \text{n.n. of }{\bf x}}}
  \tau_{\bf x} \tau_{{\bf x}'}\;, \qquad O^-_{\bf x}:= 
  \sum_{\substack{{\bf x}'\\
    \text{n.n. of }{\bf x}}} \sigma_{\bf x} \sigma_{{\bf x}'} -
  \sum_{\substack{{\bf x}'\\ \text{n.n. of }{\bf x}}} \tau_{\bf x} \tau_{{\bf
      x}'}\;.
\end{equation*}
Then, if $|K|/|J|$ is small enough, there exists one ($J$ and $K$
dependent) critical temperature at which, for large separation $|{\bf
  x}|$,
\begin{equation*}
  \langle O^+_{\bf x} O^+_0\rangle= \frac{C_+}{|{\bf x}|^{2\kappa_+}}+o(1)\;,
  \qquad \langle O^-_{\bf x} O^-_0\rangle = \frac{C_-}{|{\bf
      x}|^{2\kappa_-}}+o(1)
\end{equation*}
for $\kappa_+\equiv\kappa_+(\lambda, v)=1+O(\lambda)$,
$\kappa_-\equiv\kappa_-(\lambda, v)=1+O(\lambda)$ and
$\lambda=\frac{K}{J}$. From a mathematical viewpoint this result is
very interesting, because there was no result at all for
correlations of any of the DIM.  However,  is this result physically
significant? As opposed to the case of the Ising model, $\kappa_+$ and
$\kappa_-$, that are macroscopic quantities, do depend upon $\lambda$
and $v({\bf x})$, the parameters  that appear in the definition of the 
model. This means that the DIM class is \emph{non-universal}. However,
Kadanoff discovered that a weak form of universality still persists:
on the basis of the fact that the scaling limit of this class of
models turns out to be the Thirring model, he predicted the
formula\cite{Kad77}
\begin{equation*}
  \kappa_+(\lambda, v)\cdot \kappa_-(\lambda, v)=1\;.
\end{equation*}
Kadanoff's formula is now proven\cite{BFM09a}. The idea of the
proof is that the regularized path-integral formulation of the
Thirring model and the fermion phrasing of the partition function of
the DIM differ by irrelevant terms. Since large distances make
 irrelevant interactions negligible, as $\kappa_+\cdot \kappa_-=1$
is satisfied in the exact solution of the massless Thirring model, it
has to hold also in the lattice model. 

Let me emphasize that, in this chain of implications, 
the only knowledge of Hagen's and Klaiber's formulas, 
even if 
obtained with a rigorous limiting procedure from the Luttinger model exact solution, 
would not
suffice to deal with the scaling limit of the lattice models: 
one really needs a mathematical derivation of the Schwinger
functions from the regularized path-integral formula.

There are several other cases in which the IFP is, or could be,
resolutive. In \cite{BFM09,BFM10, BFM11} we proved similar formulas,
called ``Luttinger Liquid Relations'', for the XYZ quantum chain and
for a generalization of the $(1+1)$-dimensional Hubbard model. In
prospective, I think that the study of  critical exponents of local
bulk observables of \emph{weakly interacting dimers} (on square or
hexagonal lattices, for instance) and of the \emph{six vertex model}
(close enough to the free fermion point) should be feasible by the IFP: 
the basic calculation for the former model is showed in\cite{Fa13}. 

\section{Conclusions}
What I discussed so far is only one side of the general picture that
physicists  discovered in  the 70's and 80's, and which is 
represented in the following diagram.  
\vskip2em
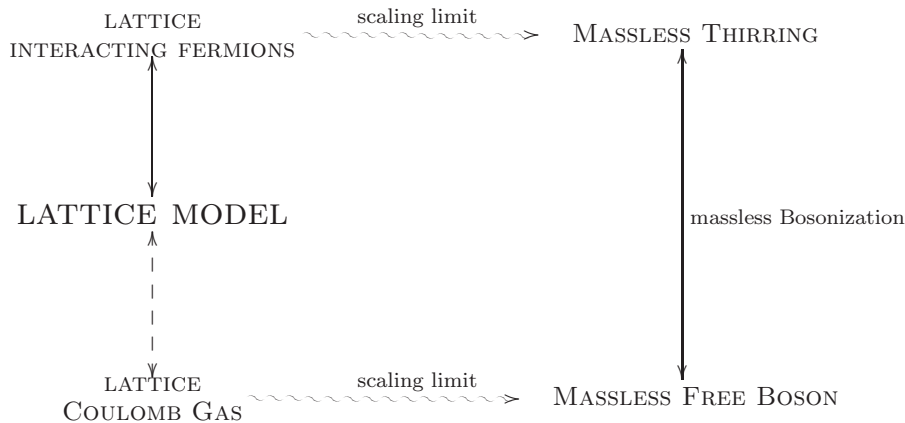
 \begin{equation*}
   \xymatrix{
     *\txt{\sc lattice\\ \sc  interacting fermions} \ar@1{~>}[rrr]^{\text{\footnotesize
         scaling limit}}\quad
     &&&\quad\text{\sc Massless Thirring}\ar@1{<->}[dddd]^{\text{\footnotesize
         massless Bosonization}}& \\ &&&\\
     \text{\LARGE\sc lattice model}\ar@1{<->}[uu] \ar@1{<-->}[dd] 
     &&&\\&&&\\
     *\txt{\sc lattice \\\sc Coulomb Gas\\} \ar@1{~>}[rrr]^{\text{\footnotesize
         scaling limit}}\quad
     &&&\quad\text{\sc Massless Free Boson}  }
 \end{equation*}
\vskip2em
The task is to compute critical exponents of lattice models
(center-right of the diagram): classical two-dimensional lattice models, 
as well as quantum one dimensional lattice models with imaginary-time dependent 
operators (i.e. ``$1+1$ dimensional models''). To achieve that, it is useful to
compute the scaling limit (i.e. the continuum limit) of such models,
which has more chances of being exactly solvable.  There are to ways
of doing that. The one explained is this review is to use the IFP, the
scaling limit of which is the Thirring model (upper part of the
diagram).  However, another approach is  possible.  I have no space for
details, but basically it consists in re-casting the lattice model
into a \emph{lattice Coulomb gas}, the scaling limit of which is the
free boson field. The agreement of the  critical exponents 
computed in the two different approaches is
explained by the bosonization of the massless Thirring model. 
The upper route, proposed by many,
including Kadanoff\cite{Kad77} and den Nijs\cite{dN79}, has been made
mathematically rigorous in some models: n.n. Ising with
n.n.n. perturbation\cite{PS00}, the XYZ quantum chain\cite{BFM09a}, a
generalization of the Hubbard model\cite{BFM11}, the class of
DIM\cite{BFM09a} and  the weakly
interacting dimers\cite{Fa13}; perhaps it is also applicable 
to the six vertex model.  
The lower route was
introduced by Kadanoff\cite{Kad78}, Nienhuis\cite{Nie84} and
others to compute critical exponents of many critical models, including: 
the $q-$states Potts model for $0\le q\le 4$;  the $O(n)$ loop model 
for $-2\le n\le 2$. Some properties of the lattice Coulomb gas are now
proved\cite{Fa12}; however,  
except for some initial progress made in \cite{FrSp81b},  
a rigorous implementation of the equivalence 
lattice model / Coulomb gas (i.e. the  broken line 
in the diagram)
is still missing.

%\bibliographystyle{ws-procs975x65} 
%\bibliography{falco}

% included bbl

\end{document}